\documentclass{PoS}

\title{One flavor mass reweighting: foundations}

\ShortTitle{One flavor mass reweighting: foundations}

\author{\speaker{Bj\"orn Leder}$^{a,b}$, Jacob Finkenrath$^a$ and
       Francesco Knechtli$^a$\\
       \llap{$^a$}Department of Physics, Bergische Universit\"at Wuppertal\\
                  Gaussstr. 20, D-42119 Wuppertal, Germany\\
       \llap{$^b$}Department of Mathematics, Bergische Universit\"at Wuppertal\\
                  Gaussstr. 20, D-42119 Wuppertal, Germany\\
       E-mail: \email{leder@physik.uni-wuppertal.de}
}

\abstract{Reweighting is not a new method in lattice QCD, but a comprehensive
analysis is missing in the literature. We close this gap by presenting: (i) a
proof of an integral representation of the complex determinant of a complex
matrix, (ii) a method to control the stochastic error of its Monte Carlo
estimation, (iii) expansions of the stochastic error and the ensemble
fluctuations of the one flavor reweighting factor. Based on (iii) we present a
detailed scaling analysis and optimized reweighting strategies.
As an application we analyze the ensemble fluctuations of the reweighting factor
corresponding to the sea contribution to isospin splitting and predict at 
physical quark masses a standard deviation of $\pm 20\%$.
}

\FullConference{31st International Symposium on Lattice Field Theory - LATTICE 2013\\
		July 29 - August 3, 2013\\
		Mainz, Germany}

\usepackage{booktabs}
\usepackage{amsmath}
\usepackage{amsthm}

\newcommand{\ev}[1]{\left\langle #1 \right\rangle}
\newcommand{\Nf}{N_{\mathrm{f}}}
\newcommand{\fm}{\,\mathrm{fm}}
\newcommand{\MeV}{\,\mathrm{MeV}}
\newcommand{\mps}{m_\mathrm{PS}}

\newcommand{\obs}{\mathcal{O}}
\newcommand{\rd}{\mathrm{d}}
\newcommand{\rD}[1]{\mathrm{D}[#1]}
\newcommand{\rO}{\mathrm{O}}
\newcommand{\e}{\mathrm{e}}

\newcommand{\mb}{{\overline{m}}}

\newcommand{\set}[1]{\{#1\}}
\newcommand{\etad}{\eta^\dagger}

\newcommand{\C}{\mathbb{C}}
\newcommand{\re}{\mathrm{Re}}
\newcommand{\im}{\mathrm{Im}}

\newcommand{\Neta}{N_{\eta}}

\newcommand{\Dm}{\Delta m}

\newcommand{\Tr}{\mathrm{Tr}}

\newcommand{\mup}{\mb_{\rm u}}
\newcommand{\md}{\mb_{\rm d}}

\newcommand{\f}{\mathrm{1f}}
\newcommand{\ff}{\mathrm{2f}}

\newcommand{\var}{\mathrm{var}}

\newcommand{\eveta}[1]{\left\langle\left\langle #1 \right\rangle\right\rangle}

\begin{document}
\vspace{-1em}
\section{Introduction}
\vspace{-1em}
In lattice QCD observables are computed as expectation values $\ev{\obs}$ of
composite fields $\obs$. The brackets stand for the average over an ensemble
of gauge configurations that is generated according to a probability distribution
$P$. Since $P$ depends on all the bare parameters of QCD a new ensemble has to be generated
for each set of bare parameters, e.g., to study the continuum limit or the mass
dependence of observables. However, it is also possible to obtain $\ev{\obs}_{P'}$ if
$P'=P\cdot W$, because then
\begin{equation}
   \ev{\obs}_{P'} = \ev{\obs W}/\ev{W}\,.
\end{equation}
The variance of the reweighting factor $W$ limits the applicability of reweighting. 
If its fluctuations become too large a reliable error estimation is impossible:
only a few configurations out of a finite ensemble of configurations from a
Monte-Carlo simulation will dominate the mean, i.e., the so-called overlap problem emerges.

Assuming $W=1/\det(I+\epsilon D^{-1}X)$ with some complex matrices $D$, $X$ and real scalar $\epsilon$,
which is the case for mass reweighting, we can expand the variance $\var(x)=\ev{x^2}-\ev{x}^2$
\begin{equation}\label{eq:fluc-exp}
   \sigma^2_W \equiv \var(W)/\ev{W}^2 = \epsilon^2 \var(\Tr(D^{-1}X)) + \rO(\epsilon^3)\,.
\end{equation}
We will come back to this expression in Section \ref{s:mrw}. But first we need to specify
how to numerically evaluate the reweighting factor $W$, i.e., the determinant of a complex matrix.
Direct computation is impossible since $D$ will turn out to be the lattice Dirac operator.
Instead we write in Section \ref{s:integral} the determinant as an integral and briefly describe
a proof for its existence for general complex matrices $A$ if $A+A^\dagger$ is positive
definite. With this proof at hand we define an unbiased and
robust stochastic
estimation of the integral in Section \ref{s:estimator}. The application of these results
to mass reweighting and numerical results for the scaling of the fluctuations are presented
in Sections \ref{s:mrw}-\ref{s:isospin}.

\vspace{-1em}
\section{Integral representation of the determinant}
\vspace{-1em}
\label{s:integral}
Let $A\in\C^{n\times n}$ and $\eta\in\C^n$ and 
$\sigma(A)$ the eigenvalues of $A$. Then one can show
\begin{equation}\label{eq:integral}
 \frac{1}{\det A} = \int \rD{\eta}\; \e^{-\etad A \eta}
   \;\quad \text{iff}\quad\; \lambda > 0\,,\; \forall \lambda\in\sigma(A+A^\dagger)
\end{equation}
with $\eta=\eta_1 e_1 + \dots + \eta_n e_n$, $\eta_i \in \C$,
$\{e_1,\dots,e_n\}$ an orthonormal basis of $\C^n$ and
\begin{equation}
   \rD{\eta} = \prod_{i=1}^n\,\frac{\rd\re(\eta_i)\,\rd\im(\eta_i)}{\pi}\,.
\end{equation}

For normal matrices the proof is straightforward since then $A$ is diagonalizable
by a unitary matrix, i.e., $A=U^\dagger \Lambda U$ with $U^\dagger U=I$ and
 $\Lambda=\mathrm{diag}(\lambda_1,\dots,\lambda_n)$. The transformation 
$\eta\to U^\dagger \eta$ has a Jacobian determinant of one and simplifies the 
multi-dimensional integral to a product of two-dimensional integrals
\begin{equation}\label{eq:normalA}
  \int \rD{\eta}\; \e^{-\etad A \eta} = \prod_{i=1}^n\,\int \frac{\rd x_i \rd y_i}{\pi}\; \e^{-\lambda_i(x_i^2+y_i^2)}\quad \text{iff $A$ normal}\,.
\end{equation}
The latter is defined iff $\re(\lambda_i)>0$ and can be solved via polar coordinates
giving $1/\lambda_i$. Since $A=U^\dagger \Lambda U$ we have 
$\sigma(A+A^\dagger)=\sigma(2\re(\Lambda))$, thus proving Eq.~\eqref{eq:integral}.

For non-normal $A$, e.g.~the Wilson Dirac operator or functions thereof, there is
no unitary transformation that diagonalizes $A$. In \cite{Finkenrath:2013soa} a proof was given
based on the Schur decomposition of $A=Q^{\dagger}(\Lambda+K)Q$, with unitary 
$Q$ and strictly upper tridiagonal $K$. Surely we have 
$\det(A)=\det(\Lambda+K)=\det(\Lambda)=\det(U^\dagger \Lambda U)$ for all unitary $U$,
i.e., the determinant of a non-normal matrix $A$ is equal to the determinant of a normal
matrix with the same eigenvalues. The proof shows that if the integrand is absolute
convergent, i.e., if $\lambda\in\sigma(A+A^\dagger)>0$, the integral representation of
$1/\det(\Lambda+K)$ is equal to that of $1/\det(\Lambda)$ by variable substitution. 

\vspace{-1em}
\section{Stochastic estimator}
\label{s:estimator}
\vspace{-1em}
If the integral representation of the determinant Eq.~\eqref{eq:integral} exists an unbiased
stochastic estimator of $W(A)=1/\det(A)$ is given by
\begin{equation}\label{eq:estimator}
   W_{\Neta}(A) = \frac{1}{\Neta}\sum_{k=1}^{\Neta} e^{-{\eta^{(k)}}^\dagger (A-I) \eta^{(k)}}\,,
\end{equation}
with Gaussian distributed\footnote{Other choices are possible. See \cite{Finkenrath:2013soa} for a
more general expression.} random vectors $\set{\eta^{(1)},\dots,\eta^{(\Neta)}}$. By the rules
of Monte-Carlo integration $W_{\Neta}$ differs from $W$ by terms of $\rO(1/\sqrt{\Neta})$.
Since this estimator can be complex we define the variance as 
$\sigma_\eta^2(A) = \eveta{W_{\Neta}(A) W_{\Neta}(A)^*} - \eveta{W_{\Neta}(A)} \eveta{W_{\Neta}(A)}^*$,
where $\eveta{O}=\int\rD{\eta}\exp(-\etad \eta)O$. It is explicitly given by the
integral representation of
\begin{equation}\label{eq:variance}
   \sigma_\eta^2(A) = \frac{1}{\det(A+A^\dagger-1)} - \frac{1}{\det(AA^\dagger)}\,.
\end{equation}
Therefore the variance exists \emph{iff} $\sigma(A+A^\dagger)>1$
(see \cite{Finkenrath:2013soa} for a proof). 
If the matrix $A$
can be written as $A=I+\epsilon B$ with $\epsilon>0$ and $\epsilon ||B||\ll 1$ the relative 
error $\delta_\eta^2=\sigma_\eta^2/(\Neta|W|^2)$ can be expanded as
\begin{equation}\label{eq:expansion}
   \delta_\eta^2(A) = \frac{1}{\Neta}[\epsilon^2 \Tr(BB^\dagger) + \rO(\epsilon^3)]\,.
\end{equation}

For the validity of the estimator in Eq.~\eqref{eq:estimator} it is enough to ensure 
$\sigma(A+A^\dagger)>0$, since then formally $W_{\Neta}\to W$ for
$\Neta\to\infty$. However, in numerical evaluations where only a small finite number of random
vectors is affordable, a well defined and controlled error with an expansion as in
\eqref{eq:expansion} is indispensable. An unbiased stochastic estimator that fulfills these
conditions automatically can be based on a factorization of $A$ (and thus $1/\det(A)$). Assume
$A=I+\epsilon D^{-1}X$ with $\epsilon||D^{-1}X||\gtrsim 1$. Then for $N\geq 1$
\begin{equation}\label{eq:factor}
   A=\prod_{i=0}^{N-1}\,[I+(\delta_{i+1}-\delta_i)D_i^{-1}X]\quad\text{with}\quad D_i=D+\delta_i X\,, \quad \delta_0=0\,,\quad \delta_{N}=\epsilon\,,
\end{equation}
if $D_i$ is invertible for all $i=0,\dots,N-1$. Note that $A$ can be written as a ratio $A=D_N/D_0$
and Eq.~\eqref{eq:factor} as a product of ratios $A=\prod_i D_{i+1}/D_i$.
The determinant $W(A)$ factorizes in the same way and the unbiased estimator is given by 
\begin{equation}\label{eq:estimator2}
   W_{\Neta,N}(A) = \prod_{i=0}^{N-1} W_{\Neta}(I+\bar{\epsilon}_i D_i^{-1}X)\,, \quad \bar{\epsilon}_i=\delta_{i+1}-\delta_i\,,
\end{equation}
where each factor uses an independent set of random vectors.
In the view of the boundary conditions in Eq.~\eqref{eq:factor} and the aim of minimizing
the stochastic error in Eq.~\eqref{eq:expansion}, the sequence $D_0,\dots,D_{N-1}$ 
should be taken as discrete steps along a smooth interpolation between $D_0$ and $D_{N-1}$
with $\bar{\epsilon}_i\to 0$ for $N\to \infty$. In the simplest case, which we assume here,
it is a linear interpolation and $\bar{\epsilon}_i \equiv \epsilon/N$. Since the eigenvalues
of $I+\bar{\epsilon}_i D_i^{-1}X$ lie in the complex plane within a circle around one whose 
radius is shrinking to zero as $N\to\infty$, the condition for the variance Eq.~\eqref{eq:variance}
to exist and the expansion in Eq.~\eqref{eq:expansion} to converge for each factor
in Eq.~\eqref{eq:estimator2} can always be fulfilled by choosing a large enough $N$. 
Because each factor is an independent estimator, the overall relative error is given by
simple error propagation and we obtain
\begin{equation}\label{eq:expansion2}
   \delta_\eta^2(A) = \frac{1}{\Neta}\sum_{i=0}^{N-1}[\bar{\epsilon}_i^2 \Tr((D_i D_i^\dagger)^{-1} X X^\dagger) + \rO(\bar{\epsilon_i}^3)] \stackrel{N\to\infty}{\leq} \frac{\epsilon^2}{N\Neta}\max_i\Tr((D_i D_i^\dagger)^{-1} X X^\dagger) \,.
\end{equation}
Thus for large enough $N$ the remaining condition for the existence of the estimator
and the variance is that $D_i$ is invertible
for all $i=0,\dots,N-1$.


\vspace{-1em}
\section{Mass reweighting in lattice QCD}
\label{s:mrw}
\vspace{-1em}
We consider mass reweighting in lattice QCD with the lattice Dirac operator $D_m=D_0+m$, where
$D_0$ is the operator at zero bare mass. Although mainly independent of the specific
fermion discretization we have $\rO(a)$-improved Wilson fermions in mind, for which we will
present some results in Sections \ref{s:oneflavor} and \ref{s:isospin}.

\vspace{-1em}
\subsection{One and two flavor reweighting}
\vspace{-1em}
The reweighting factors are determinants of ratios of Dirac operators at different
 mass parameters $m$.
We here consider one and two flavor reweighting, which covers a wide range of
applications of mass reweighting in lattice QCD. The matrix $A=I+\epsilon D^{-1}X$ 
in the general formulas above is given by
\begin{equation}\label{eq:1flavor}
   A_\f=D_{m_s-\Dm}^{-1} D_{m_s} =I+\Dm D_{m_s-\Dm}^{-1}\,,
\end{equation}
for one flavor reweighting and 
\begin{equation}\label{eq:2flavor}
   A_\ff= (D_{m_r-\gamma\Dm} D_{m_s+\Dm})^{-1}D_{m_r} D_{m_s}=I+\Dm (D_{m_s+\Dm}D_{m_r-\gamma\Dm})^{-1} (\gamma\Dm+\gamma D_{m_s}-D_{m_r})\,,
\end{equation}
for two flavor reweighting. In each case $m_{r,s}$ are the ensemble mass parameters and
$m_s\pm\Dm$ and $m_r-\gamma\Dm$, respectively, are the target mass parameters. For a 
detailed discussion of Eq.~\eqref{eq:2flavor} we refer to \cite{Finkenrath:2013soa}.
The special case $m_r=m_s$ and $\gamma=-1$, i.e., reweighting of a mass-degenerated pair of
quarks, was considered in \cite{Hasenfratz:2008fg}.
Here we concentrate on the special case $m_r=m_s$ and $\gamma=1$, which we dub
 isospin reweighting
\begin{equation}\label{eq:iso}
   A_\pm=I+\Dm^2 (D_{m_s}^2-\Dm^2)^{-1}\,.
\end{equation}

\vspace{-1em}
\subsection{Twisted mass reweighting}
\vspace{-1em}
Applying the estimator of Section \ref{s:estimator} we have to ensure that $D_m$ is
invertible for all occurring values of $m$ and for all configurations of a given ensemble.
For Wilson fermions this can in general not be guaranteed. A solution is to add a small
twisted mass $D_0\to D_0 +i\gamma_5\mu$, because then the lattice Dirac operator has a gap.
If a finite twisted mass is already included in the simulation as proposed in \cite{Luscher:2012av},
this does not mean any additional effort. If the ensemble was generated at $\mu=0$ an additional
reweighting to finite $\mu$ and back is necessary. In \cite{Finkenrath:2013soa} we proposed to do this only
for those configurations that suffered from small eigenvalues of $D_m$ for $\mu=0$. However,
to ensure the
interchangeability of the integral over gauge fields and random vectors, it is necessary to do this
for every configuration\footnote{We thank M. L\"uscher and S. Schaefer for pointing this
 out.}. The analysis of the fluctuations and stochastic estimators in 
Section \ref{s:estimator} can be applied to the case of twisted mass reweighting as well and is
currently under investigation.

\vspace{-1em}
\subsection{Fluctuations and stochastic error} 
\vspace{-1em}
Assuming $\Dm>0$ and $m=m_s-\Dm$ we obtain for the relative stochastic error and the
fluctuations
\begin{equation}\label{eq:errorvar}
   \ev{\delta_p^2} \approx \frac{\Dm^{2p}}{N\Neta}\ev{\Tr((D_m D_m^\dagger)^{-p})} \quad\text{and}\quad
   \sigma_p^2 \approx \Dm^{2p} \var(\Tr(D_m^{-p}))\,.
\end{equation}
with $p=1$ for one flavor and $p=2$ for isospin reweighting.
Some insight into the volume and (renormalized) quark mass dependence of the involved traces can in
principal be obtained from chiral perturbation theory, e.g., as in \cite{Giusti:2008vb}.
For example, at lowest order in the chiral expansion
\begin{equation}\label{eq:chpt} 
   \ev{\Tr((D_m D_m^\dagger)^{-p})} = \frac{\mb\Sigma V}{\mb^{2p}} \frac{\Gamma(p-\frac{1}{2})}{\sqrt{\pi}\,\Gamma(p)}\,,
\end{equation}
with the volume $V$, the chiral condensate $\Sigma$ and the renormalized quark mass $\mb$.
In lack of an explicit calculation for $\var(\Tr(D_m^{-p}))$ we
assume the ad-hoc scaling formula $k\cdot (V/a^4)/(a^{q-2p}\mb^q)$
with some dimensionless constant $k$ and some power $q$ of the renormalized quark mass.

\vspace{-1em}
\section{One flavor reweighting}
\label{s:oneflavor}
\vspace{-1em}
In \cite{Finkenrath:2013soa} numerical results where presented for mass reweighting
of one ensemble, tagged D5, of
$\Nf=2$ $\rO(a)$-improved Wilson fermions at a lattice spacing of $a=0.066\fm$ \cite{Fritzsch:2012wq}. Here we add a second
ensemble, tagged F7, at a smaller pion mass $\mps$ and with lager volume to the
analysis. The ensemble was generated within the CLS
effort\footnote{https://twiki.cern.ch/twiki/bin/view/CLS/}. We list the lattice volumes,
the pion masses and the renormalized quark masses, as defined in
 \cite{Fritzsch:2012wq}, for
the two ensembles in Tab.~\ref{tab:ensembles}. We also give the maximal renormalized reweighting
distance, where  $\Delta \mb$ is the difference of the renormalized quark masses before and after reweighting
$\Delta \mb=\mb(m)-\mb(m-\Dm)$.

\begin{table}[tp]
 \centering
\begin{tabular}{ccccc}
\toprule
  & $V/a^4$ & $\mps$ in $\MeV$ & $\mb$ in $\MeV$ & $\max(|\Delta \mb|)$ in $\MeV$\\
\midrule
D5 & $48\times 24^3$ & $440$ & $32$ & $3$ \\
F7 & $96\times 48^3$ & $270$ & $12$ & $3$ \\
\bottomrule
\end{tabular}
 \caption{Ensembles and maximal reweighting distance considered for the numerical results.
 The pion masses $\mps$ and the renormalized quark masses $\mb$ are rounded to steps
 of $5 \MeV$ and $1 \MeV$, respectively.}
 \label{tab:ensembles}
\end{table}

\begin{figure}[htp]
 \centering
 \includegraphics[width=0.43\textwidth]{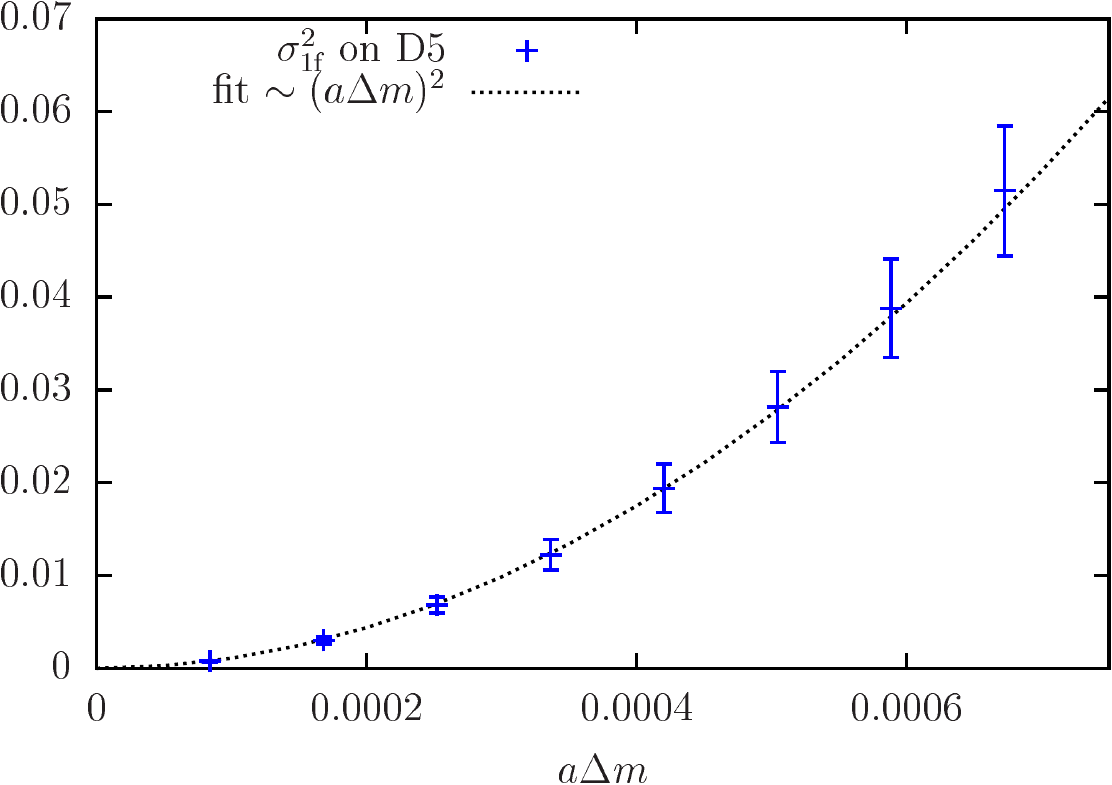}\hspace{2em}
 \includegraphics[width=0.43\textwidth]{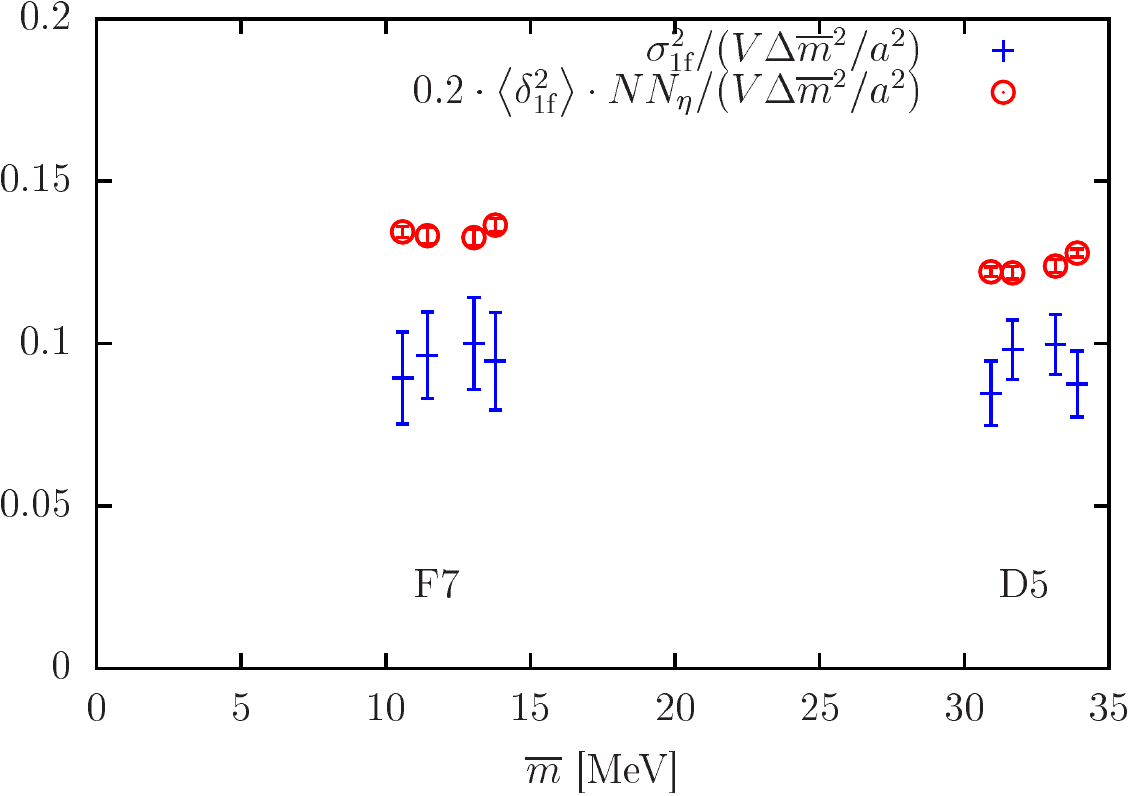}
 \caption{Left: Dependence of the ensemble fluctuations $\sigma_\f^2$ of one flavor reweighting 
 on the reweighting distance $\Dm$ in lattice units for ensemble D5. 
 Right: Dependence of $\sigma_\f^2$ and the
 stochastic error $\delta_\f^2$ on the renormalized quark mass $\mb$. Both quantities are
 normalized such that differences in the volume 
 and the renormalized reweighting distance $\Delta \mb$ are removed. 
 The stochastic error $\ev{\delta_\f^2}$ is multiplied by $0.2$ for visibility. Out of the 8 points
 in the left panel we plot the one with the largest $a\Dm$, the one at half this distance and
 two points that correspond to the same absolute value but the opposite sign of $\Dm$.
The number of random vectors vectors is fixed to $\Neta=6$ and $N=8$ at (or $N=4$ at half)
the maximal reweighting distance.
 The renormalized quark mass is the average mass of the two quarks after reweighting.}
 \label{fig:oneflavor}
\end{figure}

For the reweighting distances considered here the expansions in Eqs.~\eqref{eq:fluc-exp} and
\eqref{eq:expansion2} work and are dominated by the first term. We demonstrate this in the
left panel of Fig.~\ref{fig:oneflavor} for the ensemble fluctuations. The behavior of
the stochastic error is similar. The two ensembles at different quark masses allow for
a study of the quark mass dependence of the reweighting factor. As explained in Section
\ref{s:mrw} we expect for one flavor reweighting
\begin{equation}\label{eq:errorvar1}
   \ev{\delta_\f^2} \approx \frac{k_{\eta,\f}}{N\Neta}\frac{\Delta \mb^{2} V}{a^{(2+q)} \mb^q}\quad\text{and}\quad
   \sigma_\f^2 \approx k_\f\frac{\Delta \mb^{2} V}{a^{(2+q')}\mb^{q'}} \,,
\end{equation}
where we replaced $\Dm$ by the difference of the renormalized masses. Lowest order
chiral perturbation theory predicts $q=1$. In order to reveal
the mass dependence we plot in the right panel of Fig.~\ref{fig:oneflavor}
$\sigma^2_{\text{1f}}/(V\Delta\overline{m}^2/a^2)$ and
$\ev{\delta^2_{\text{1f}}}\cdot N\Neta /(V\Delta\overline{m}^2/a^2)$ as function of the (average)
renormalized quark mass after reweighting.
The four points for each ensemble correspond to $\pm \max(|\Delta \mb|)$ and half this
distance. Note that we neglect errors on the $x$-axis.

There is no visible dependence on the quark mass, so one would conclude $q=q'=0$.\
This result is somewhat puzzling,
since for very large mass the fluctuations should go to zero and for very small mass
the chiral expansion should describe the data. These results seem to be obtained
in an intermediate regime.

\vspace{-1em}
\section{Isospin reweighting}
\label{s:isospin}
\vspace{-1em}

\begin{figure}[htp]
 \centering
 \includegraphics[width=0.43\textwidth]{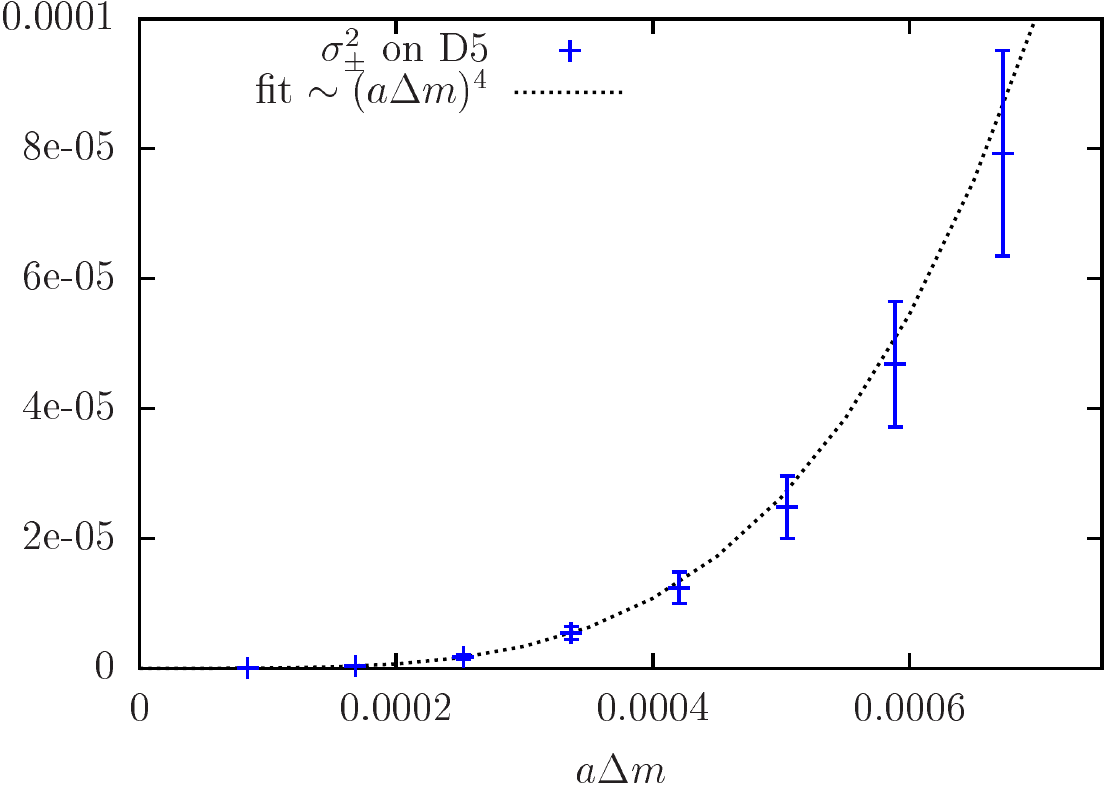}\hspace{2em}
 \includegraphics[width=0.43\textwidth]{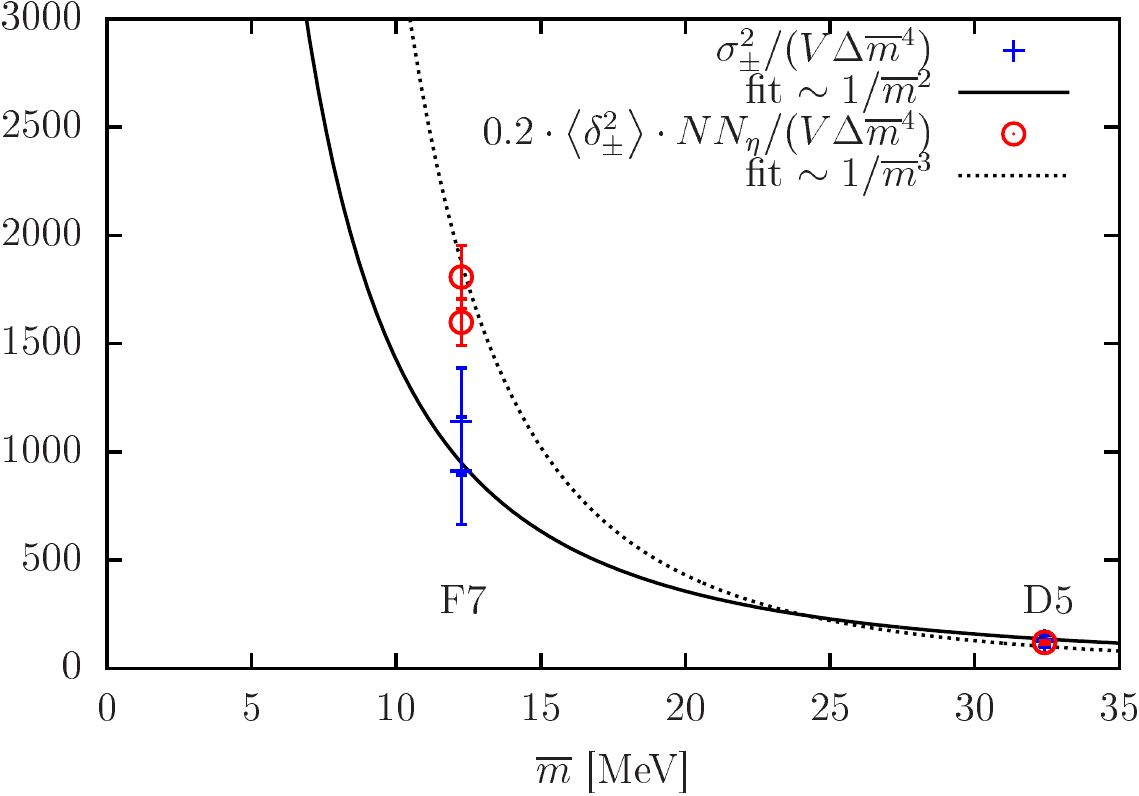}
 \caption{Same as Fig.~\protect\ref{fig:oneflavor} but for isospin reweighting, i.e.,
  $\sigma_\pm^2$ and $\ev{\delta_\pm^2}$.}
 \label{fig:isospin}
\end{figure}

In the case of isospin reweighting the two quarks are reweighted simultaneously in opposite
direction, as in Eq.~\eqref{eq:iso}. The expansions in Eqs.~\eqref{eq:fluc-exp} and
\eqref{eq:expansion2} work also in this case. We demonstrate this in the
left panel of Fig.~\ref{fig:isospin} for the ensemble fluctuations. Again the behavior of
the stochastic error is similar. Therefore we expect for isospin reweighting
\begin{equation}\label{eq:errorvar2}
   \ev{\delta_\pm^2} \approx \frac{k_{\eta,\pm}}{N\Neta}\frac{\Delta \mb^{4} V}{a^{q} \mb^q}\quad\text{and}\quad
   \sigma_\pm^2 \approx k_\pm\frac{\Delta \mb^{4} V}{a^{q'}\mb^{q'}} \,,
\end{equation}
with $q=3$ indicated by lowest order chiral perturbation theory.
 In the right panel of Fig.~\ref{fig:isospin} we plot
$\sigma^2_{\pm}/(V\Delta\overline{m}^4)$ and
$\ev{\delta^2_{\pm}}\cdot N\Neta/(V\Delta\overline{m}^4)$ as function of the renormalized quark mass 
after reweighting.
The two points for each ensemble correspond to $\max(|\Delta \mb|)$ and half this
distance. There clearly is a strong dependence on the mass and we also plot the result of
one parameter fits of the fluctuations and the stochastic error with $q'=2$ and $q=3$.
If we assume Eq.~\eqref{eq:chpt} to be applicable, i.e., higher order terms and finite
volume effects to be negligible, the latter fit yields a prediction for the chiral condensate
$\Sigma=(325 \MeV)^3$.

\vspace{-1em}
\section{Conclusion}
\vspace{-1em}
Reweighting factors in lattice QCD almost always include determinants of ratios
of the lattice Dirac operator, i.e., large sparse matrices. The numerical evaluation
is necessarily stochastic and based on an integral representation of the determinant.
Stochastic estimators with controlled variance can be defined if the determinant
is factorized.

Both, the stochastic error of the reweighting factor and its 
ensemble fluctuations can be expanded in the reweighting distance. For small
reweighting distances they are dominated by the first term in this expansion.
In the case of one flavor and isospin reweighting the dependence on the renormalized
quark mass is analyzed. Whereas for one flavor reweighting there is no dependence in
the range considered, for isospin reweighting a strong dependence, $\propto 1/\mb^{2}$
for the ensemble fluctuations, is found. Although not significant at the quark masses
considered here, they will become important at physical quark masses. From the value
for F7 in the right panel of Fig.~\ref{fig:isospin}, the numbers in
Tab.~\ref{tab:ensembles} and $\Delta \mb=(\md-\mup)/2\approx1.5 \MeV$ one obtains $\sigma^2_{\pm}(\text{F7})\approx 0.0007$. Compared
to F7 physical
masses mean a factor four smaller quark mass and a factor $(4/3)^4$ larger volume to
keep finite volume effects small. Thus  $\sigma^2_{\pm}(\text{phys. mass})\approx 0.034$,
which can have a significant effect on observables.

\vspace{-1em}

\end{document}